\documentstyle[11pt,newpasp,twoside,epsf]{article}
\def\edcomment#1{\iffalse\marginpar{\raggedright\sl#1\/}\else\relax\fi}
\reversemarginpar

\begin{document}
\title{Infrared Properties of Cataclysmic Variables from 2MASS: Results 
from the 2nd Incremental Data Release}

\author{D. W. Hoard and S. Wachter} 
\affil{Cerro Tololo Inter-American Observatory, Casilla 603, La Serena, Chile}

\author{L. L. Clark} 
\affil{Dept.\ of Astronomy, San Diego State University, San Diego CA 92182}

\author{Timothy P. Bowers} 
\affil{Steward Observatory, University of Arizona, Tucson AZ 85721}

\begin{abstract}
Because accretion-generated luminosity dominates the radiated energy of 
most cataclysmic variables (CVs), they have been ``traditionally'' 
observed primarily at short wavelengths.  
Infrared (IR) observations of CVs contribute to the understanding of 
key system components that are expected to radiate at these 
wavelengths, such as the cool outer disk, accretion stream, and 
secondary star.
We have compiled the $J$, $H$, and $K_{\rm s}$ photometry of all 
CVs located in the sky coverage of the 2 Micron All Sky Survey (2MASS) 
2nd Incremental Data Release.  
This data comprises 251 CVs with reliably identified near-IR 
counterparts and S/N$>$10 photometry in one or more of the three 
near-IR bands.
\end{abstract}

\section{Orbital Period-Color Diagram}

One of the hoped for goals of observing CVs in the IR is to isolate the 
luminosity contribution of the secondary stars, and learn more about 
the mass donors in these systems.
Figure 1 shows the $(J-K{\rm s})$ colors of the CVs in the 2MASS 2nd 
Incremental Data Release Point Source Catalog for systems with known 
orbital period (as compiled in Downes et al.\ 2001).
The thick curve shows the color of the expected CV secondary star 
as a function of orbital period (Smith \& Dhillon 1998).
The horizontal bars show the range of orbital period over which each 
secondary star spectral type is found.
If the near-IR data truly isolated the secondary star luminosity, 
then all of the points would lie along the curve inside the range 
delimited by the bars.  For the (admittedly few) long period 
systems ($P_{\rm orb}>7.5$ hr), this expectation is met fairly well.
However, the majority of short orbital period CVs are offset blueward of 
the $(J-K_{\rm s})$ color of their expected secondary stars.  
This blue contamination of their near-IR luminosities is almost certainly 
related to the accretion process.

Four CVs are labeled in the figure; these are U Gem, SS Cyg, RW Tri, 
and TV Col.  
These systems have the best known distances of all CVs, determined 
from trigonometric parallaxes measured with the {\em HST} Fine 
Guidance Sensors (Harrison et al.\ 1999; McArthur et al.\ 1999, 2001).
All four of these CVs fall remarkably close to the main sequence when 
their distances are used to place them in an $M_{K}$ vs.\ $(J-K)$ 
color-magnitude diagram.  
However, Figure 1 clearly shows that only U Gem has a near-IR color 
consistent with that of the secondary star expected at its orbital period.  
The other three CVs are substantially contaminated by blue light from 
the accretion process.

\begin{figure}[bth]
\plotfiddle{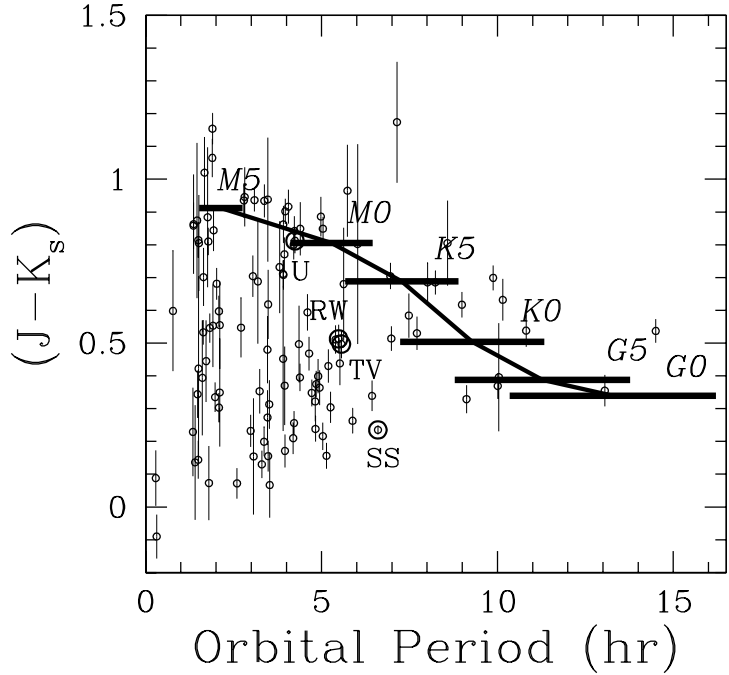}{3.35in}{0}{49}{49}{-150}{-75}
\caption{IR color of CVs as a function of orbital period.}
\end{figure}

For more information about the IR properties of CVs from 2MASS, see
{\tt http://www.ctio.noao.edu/$\sim$hoard/research/2mass/}
and Hoard et al.\ (2001).

\bigskip
\acknowledgements{This publication makes use of data products from 
the 2 Micron All Sky
Survey, which is a joint project of U.\ of Massachusetts and 
IPAC/CalTech,
funded by NASA and the NSF.}

\end{document}